\documentclass[%
 reprint,
 superscriptaddress,
 nofootinbib,
 amsmath,amssymb,
 aps,
 prl,
]{revtex4-2}

\usepackage{graphicx}
\usepackage{dcolumn}
\usepackage{bm}

\usepackage{booktabs}
\usepackage[english]{babel}
\usepackage{amsmath,amssymb,amsbsy,amstext, amsthm, simplewick}
\usepackage[hidelinks]{hyperref}
\usepackage{graphicx}
\usepackage{amsfonts}
\usepackage{amssymb}
\usepackage{upgreek}
\usepackage{cancel}
\usepackage{color}
\usepackage{slashed}
\usepackage[dvipsnames]{xcolor}
\usepackage{tikz-cd}
\usepackage{feynmp-auto}
\usepackage{color}
\usepackage{soul}
\usepackage{dsfont}
\usepackage{verbatim}
\usepackage[utf8]{inputenc} 
\usepackage{soul}
\usepackage{slashed}

\newcommand{\gen}{{N_{\text{g}}}}
\renewcommand{\t}[1]{\tilde{#1}}
\newcommand{\bb}[1]{\mathbb{#1}}

\def\@fnsymbol#1{\ensuremath{\ifcase#1\or $\Re$\or $\Im$\or  \else\@ctrerr\fi}}

\begin{document}

\title{A Note on Proton Stability in the Standard Model}
\author{Seth Koren}
\affiliation{Enrico Fermi Institute, University of Chicago,
Chicago, IL 60637, U.S.A.}

\begin{abstract}
In this short note we describe the symmetry responsible for absolute, nonperturbative proton stability in the Standard Model. 
The SM with $N_c$ colors and ${N_g}$ generations has an exact, anomaly-free, generation-independent, global symmetry group $U(1)_{B-N_c L} \times \mathbb{Z}_{N_g}^L$, which contains a subgroup of baryon \textit{plus} lepton number of order $2 N_c {N_g}$.
This disallows proton decay for ${N_g}>1$. 
Many well-studied models beyond the SM explicitly break this global symmetry, and the alternative deserves further attention.\\
\end{abstract} 
 
\maketitle

Everything not forbidden is compulsory.
So which symmetry forbids proton decay in the Standard Model? It's the lightest baryon, but baryon number is anomalous and not a symmetry of the quantum SM. The difference between baryon and lepton number is anomaly-free, but allows e.g. $p^+ \rightarrow e^+ \pi^0$. In fact there is a discrete subgroup of baryon plus lepton number which is anomaly free by virtue of the SM having more than one generation. This symmetry imposes the selection rule $\Delta B = N_c \gen, \Delta L = \gen$ on the SM with $N_c$ colors and $\gen$ generations. In the following we briefly review the topic of mixed anomalies in the SM selectively aimed toward evincing the anomaly-free discrete global symmetries. Field theoretic calculations we have omitted can be found in standard QFT textbooks or in Bertlmann's monograph \cite{Bertlmann:1996xk}. 

\begin{table}[]
\large
\begin{tabular}{|c|c|c|c|c|c|}  \hline
 & $\quad Q \quad$ & $\quad\bar u\quad$ & $\quad\bar d\quad$ & $\quad L \quad$ & $\quad\bar e\quad$ \\ \hline
$SU(3)_C$ & $3$ & $\bar 3$ & $\bar 3$ & -- & -- \\ \hline
$SU(2)_L$ & $2$ & -- & -- & $2$ & -- \\ \hline
$U(1)_{Y}$ & $+1$ & $-4$ & $+2$ & $-3$ & $+6$ \\ \hline
$U(1)_{B}$ & $+1$ & $-1$ & $-1$ & -- & -- \\ \hline
$U(1)_{L}$ & -- & -- & -- & $+1$ & $-1$ \\ \hline
\end{tabular}\caption{Representations of the SM Weyl fermions under the classical symmetries of the SM. We normalize each $U(1)$ so the least-charged particle has unit charge, $ B \equiv 3 B_{\text{usual}}$, $Y \equiv 6 Y_{\text{usual}}$, and $ L \equiv L_{\text{usual}}$.  
}\label{tab:charges}
\end{table}

The Standard Model of particle physics is defined as the gauge theory of the non-Abelian symmetry group $SU(3)_C \times SU(2)_L \times U(1)_Y$ with three `generations' (or `families') of left-handed Weyl fermions in the representations shown in Table \ref{tab:charges}. There is additionally a scalar electroweak Higgs doublet which has Yukawa couplings providing masses to the electrically-charged fermions in the broken phase. 

The SM so defined contains additional `accidental' generation-independent exact classical global symmetries corresponding to baryon and lepton number, whose charges are also listed in Table \ref{tab:charges}. These are accidental in that the most general renormalizable Lagrangian one may write down automatically preserves them. However, the Lagrangian is a classical object and a good classical global symmetry $U(1)_X$ may be broken by the path integral measure upon quantization if the fermions charged under $U(1)_X$ are in a chiral representation of a gauge group $G$ \cite{Fujikawa:1979ay}.  

One may check whether the classical global symmetry survives quantization by examining the `anomaly conditions', which in four dimensions consists essentially of evaluating the three-point correlator of the symmetry currents at one loop and checking if the Ward-Takahashi identity is satisfied. Our case of interest will have one global symmetry current and two gauge symmetry currents, such that the condition for global current conservation in the quantum theory is 
\begin{equation}
    \partial_{\mu} \langle J^{\mu}_X J^{\nu}_G J^{\rho}_G \rangle = 0.
\end{equation}
If this condition is satisfied, then the symmetry $U(1)_X$ is `anomaly-free' and the classical current conservation $\partial_{\mu} J^{\mu}_X = 0$ may be upgraded to a Ward identity in the full quantum theory. If this condition is violated then nonperturbative $G$ gauge theory effects inevitably lead to $U(1)_X$ symmetry violation, and furthermore, $U(1)_X$ cannot itself be consistently gauged.\footnote{The absence of ABJ anomalies is necessary to gauge $U(1)_X$, but not sufficient. We note in particular that the $\left\langle J_X J_X J_X\right\rangle$ `\textquotesingle t Hooft anomaly' need also vanish, and in the presence of dynamical gravity, a mixed anomaly with gravitational currents must also be avoided.}

This type of anomaly is sometimes referred to as a `mixed anomaly' between $U(1)_X$ and $G$ or an `ABJ anomaly' after its discovery by Adler-Bell-Jackiw \cite{Adler:1969gk,Bell:1969ts} as the microphysical explanation for the decay of the neutral pion $\pi^0 \rightarrow \gamma \gamma$. In that application one relates the failure of the Ward identity in the three-point correlator to a three-point amplitude using the LSZ formula. The pion, as a pseudo-Goldstone of the axial symmetry $U(1)_A$, has nonzero overlap with its global symmetry current $J_A^\mu$, and the photons have nonzero overlap with the $U(1)_Q$ gauge symmetry current $J_Q^\mu$, so this amplitude provides for pion decay.

When such an anomaly is present, a global rotation of the charged fermions does not leave the action invariant but rather changes the effective $\theta$-term for the $G$ gauge field. If we perform a rotation of the fermions charged under the global $U(1)_X$ by an angle $\alpha$, the action is shifted as
\begin{equation} \label{eq:shift}
\psi_i \rightarrow \psi_i e^{i q_i \alpha} \quad \Rightarrow \quad \delta S = \alpha \mathcal{A} \int \frac{F \tilde{F}}{16\pi^2},
\end{equation}
where $\psi_i$ are left-handed Weyl fermions with charge $q_i$ under $U(1)_X$, $F$ is the field strength of the gauge group $G$ and $\tilde{F}$ is its Hodge dual. If this transformation changes the partition function of the quantum theory, then it is no longer a symmetry. 

The integrand may be recognized as the Chern-Pontryagin density of the gauge field configuration, and its integral is an integer topological invariant which measures the winding of the gauge field around the sphere at Euclidean spacetime infinity. $\mathcal{A}$ is also an integer which is the sum of the `anomaly coefficients' of all left-handed Weyl fermions
\begin{equation}
\mathcal{A} \ \delta^{ab} = \sum\limits_{i} \text{Tr} \left[q_i T^a_{R_i} T^b_{R_i}\right],
\end{equation}
where $T^a_{R_i}$ are the generators of the representation $R_i$ of $G$ and for a non-Abelian group the normalization is such that in the fundamental representation $F$, $\text{Tr}\left[T_F^a T_F^b\right] = \delta^{ab}$. That is, a $G$ fundamental with unit $X$ charge contributes $\mathcal{A} = 1$. For an Abelian $G$ there is only one generator and $\delta^{ab} \mapsto 1$. The integral nature of the anomaly derives from the number of zero modes of fermions in the background spacetime---these are counted by the index of the Dirac operator, which is directly related to the anomaly by a theorem of Atiyah and Singer \cite{atiyah1963index,Fujikawa:1980eg}. 

If $G$ is non-Abelian and the only nontrivial representations of $G$ are fundamentals, as for example with $SU(2)_L$ in the SM, then we have simply
\begin{equation}
\mathcal{A}_{X SU(2)^2_L} = \sum\limits_{\text{Fund} \ i} q_i,
\end{equation}
while if $G$ is abelian as for example with $U(1)_Y$ in the SM, we have simply a trilinear in charges
\begin{equation}
\mathcal{A}_{X U(1)_Y^2} = \sum\limits_{i} q_i Y_i^2.
\end{equation}

If the fermions charged under $U(1)_X$ are in a vector-like representation of $G$, as all the SM fermions are with $SU(3)_C$ (or electromagnetism), then for each left-handed Weyl fermion $\psi_i$ there is another right-handed Weyl fermion $\psi_{\bar i}^\dagger$ with the same quantum numbers, such that they pair up into a Dirac spinor. Then $\psi_i$ and $\psi_{\bar i}$ are in complex conjugate representations of the gauge group and their contributions to the anomaly coefficient are related by three negative signs and cancel out.

\begin{table}[h]
\large
\begin{tabular}{|c|c|c|}  \hline
 & $U(1)_{B}$ & $U(1)_{L}$ \\ \hline
$SU(2)_L^2$ & $N_c \gen$ & $\gen$ \\ \hline
$U(1)_{Y}^2$ & $-18 N_c \gen$ & $- 18 \gen$ \\ \hline
\end{tabular}\caption{Mixed anomalies of the classical accidental symmetries with the chiral gauge symmetries of the SM. $N_c$ is the number of colors and $N_g$ the number of generations.}\label{tab:mixedanom}
\end{table}

In Table \ref{tab:mixedanom} we give the ABJ anomalies of baryon and lepton number with the chiral factors of the SM gauge group. $\mathcal{A} \neq 0$ indicates the presence of an anomaly and the classical $U(1)_X$ symmetry is broken, since the action is no longer invariant under the transformation as in Equation \ref{eq:shift}. As is familiar, while both baryon and lepton number have anomalies, we may form the anomaly-free current $B -  N_c L$ of baryon minus lepton number. This is the only anomaly-free, generation-independent continuous global symmetry of the SM. On the other hand, baryon \textit{plus} lepton number is violated in nonperturbative processes involving the gauge fields given by the new term in the action Equation \ref{eq:shift}, which effects (see e.g. \cite{Morrissey:2005uza} for lucid discussion)
\begin{equation}\label{eq:currentnoncons}
    \left\langle \partial_\mu J^\mu_{B+N_cL} \right\rangle = 2 N_c \gen \int \frac{W \t{W}}{16 \pi^2},
\end{equation}
where the expectation value is taken in a given background gauge field, $W$ is the $SU(2)_L$ field strength in that configuration, and we have left off the similar $U(1)_Y$ term as it has no effects in $d=4$ flat space for reasons of topology. This nonperturbative effect is central to electroweak baryogenesis \cite{Kuzmin:1985mm,Shaposhnikov:1987tw} wherein the thermal configurations giving dynamical symmetry violation are called `sphalerons' \cite{Manton:1983nd,Klinkhamer:1984di}.

While we have exhausted the anomaly-free continuous global symmetries, let us now relax our symmetry of interest from the full $U(1)_X$ of rotations by arbitrary angles to the subgroup of transformations by $\alpha = 2\pi k / N$ for some $N \in \bb{N}, k=0..N-1$. If we choose $N = \mathcal{A}$, then under any rotation the action changes by a multiple of $2\pi i$ in Equation \ref{eq:shift} and the partition function is invariant. This $\bb{Z}_N$ subgroup of $U(1)_X$ then remains a good symmetry of the quantum theory. 

In the case of the SM, this means that there is an additional discrete, anomaly-free $\bb{Z}_\gen$ worth of symmetries for the SM with $\gen$ generations of fermions. Of course there is some freedom to describe the additional generator, since any addition of $Y$ or $B-N_c L$ would work just as well. But we may non-redundantly identify this as the $\bb{Z}_\gen^L$ subgroup of lepton number $U(1)_L$, which is manifestly independent of both---whereas there is a $\bb{Z}_{N_c}$ subgroup of $B-N_c L$ in which the leptons transform trivially and the transformation is equivalent to one of $U(1)_B$. 

Consequently, the anomaly-free, generation-independent, global symmetry group of the SM is $U(1)_{B-N_c L} \times \bb{Z}_\gen^L$. A few more remarks are in order on the structure of this symmetry group. Firstly, we note that baryon \textit{plus} lepton number $U(1)_{B+N_cL}$ intersects this group in a $\bb{Z}_{2N_c \gen}$ subgroup generated by $(1,1) \in \mathbb{Z}_{2 N_c \gen}^{B - N_c L} \times \mathbb{Z}_{\gen}^L$. 

This $\bb{Z}_{2N_c \gen}$ is the maximal anomaly-free subgroup of baryon plus lepton number in the SM. We see that the appearance of the anomaly coefficient in Equation \ref{eq:currentnoncons} expressing current nonconservation enforces dynamically the $\Delta L = \gen, \Delta B = N_c \gen$ selection rule imposed by the existence of this exact discrete symmetry. Indeed, SM sphaleron processes all respect this selection rule.

Further within this, we note that there is a $\bb{Z}_{2N_c}$ subgroup in which $U(1)_{B+N_cL}$ intersects $U(1)_{B-N_cL}$ directly, since leptons and antileptons have the same charge $\text{mod} \ 2 N_c$. And inside of this, fermion number can be realized as the order two subgroup of $B \pm N_c L$ rotations by $e^{i\pi F}$, since the only fields charged under $B$ or $L$ in the SM have odd $B$ or $L$ charges and are fermions, and $N_c$ is also odd. Summarizing these relationships, we have \[U(1)_{B-N_cL}\times \bb{Z}^L_\gen \supset \bb{Z}_{2N_c \gen}^{B+N_cL} \supset \bb{Z}_{2N_c}^{B\pm N_cL} \supset (-1)^F,\]
among anomaly-free global symmetries of the SM.

We note also that the $U(1)_B$ symmetry we have defined in the SM at high energies may more accurately be named `quark number'. It is in the confined phase that $B_{\text{usual}} \equiv B/N_c$ really counts baryons, which must be constructed with the $SU(N_c)$ invariant tensor $\varepsilon_{i_1 i_2 \dots i_{N_c}}$ to be colorless. If we strictly work in an effective theory below nuclear energy scales then there are no $B-N_cL$ unit charges, and it's sensible to work with baryon minus lepton number as usually defined $B/N_c - L$. The above subgroup series is then modified to
\[U(1)_{B/N_c - L}\times \bb{Z}^L_\gen \supset \bb{Z}_{2 \gen}^{B/N_c+L} \supset (-1)^F.\]
The proton, with $B/N_c = 1$ and as the lightest baryon in the broken phase, then cannot decay while satisfying both $\Delta(B/N_c - L) = 0$ and $\Delta(B/N_c + L) = 0 \ (\text{mod} \ 2 \gen)$.

While many theories beyond the SM explicitly break these global symmetries, in the face of increasingly stringent constraints on the lifetime of the proton it may be worth reconsidering the prospects that it is absolutely stable. I, for one, would welcome the possibility of there being one fewer looming existential threat.




\section*{Earlier Work}

Work on anomalies of discrete symmetries began with \cite{Ibanez:1991hv,Ibanez:1991pr}. A variety of authors have considered exotic \textit{gauged} $\bb{Z}_3$ symmetries to stabilize the proton in the context of the Minimal Supersymmetric Standard Model (MSSM) and extensions thereof (e.g. \cite{Ibanez:1992ji,Hinchliffe:1992ad,Kubo:2001cr,Dreiner:2005rd,Dreiner:2006xw,Lee:2007qx,Luhn:2007gq,Lee:2010vj,Anastasopoulos:2015bxa}), where baryon and lepton number are no longer classical global symmetries. Other interesting related work includes \cite{Banks:1991xj,Preskill:1991kd,Csaki:1997aw,Dine:2007zp,Araki:2008ek,Berasaluce-Gonzalez:2011gos,Evans:2012a,Tachikawa:2017gyf,Kobayashi:2021xfs,Anber:2021upc,Davighi:2022qgb}. I note especially that, while conducting extensive literature review, I found that \cite{Tong:2017oea} on the global structure of the SM gauge group noted the existence of a $\bb{Z}_\gen$ anomaly-free subgroup of $B/N_c + L$, and \cite{Byakti:2017igi} on discrete symmetries in the MSSM mentioned in their Footnote 7 that a $\bb{Z}_\gen$ subgroup of $B/N_c$ protects the proton in the SM. Soon after this work appeared, \cite{Wang:2022eag} explored the incompatibility of the $B/N_c + L$ symmetry with a variety of grand unification schemes. I beg the pardon of any experts who know the facts explained in this manuscript already, and I hope the preceding dedicated discussion remains of use to the community.


\section*{Acknowledgements}

I am grateful to Clay Cordova, Carlos Wagner, Liantao Wang, and especially T. Daniel Brennan and Sungwoo Hong for helpful discussions. This work was supported by an Oehme Postdoctoral Fellowship from the Enrico Fermi Institute at the University of Chicago.

\bibliography{proton}

\end{document}